\documentclass[preprint, 3p, times]{elsarticle}

\usepackage[latin1]{inputenc}
\usepackage{subfig}
\usepackage{amsfonts,amsmath,amssymb,amsthm}
\usepackage{graphicx}
\usepackage[dvips]{color}
\usepackage{bm}
\usepackage{algorithm}
\usepackage{algorithmic}
\usepackage{relsize}
\usepackage{accents}

\makeatletter
\def\ps@pprintTitle{%
   \let\@oddhead\@empty
   \let\@evenhead\@empty
   \def\@oddfoot{\reset@font\hfil\thepage\hfil}
   \let\@evenfoot\@oddfoot
}
\makeatother

\begin{document}
\begin{frontmatter}



\title{A model for an aquatic ecosystem}

\author{Han Li Qiao, Ezio Venturino}

\address{Department of Mathematics \lq\lq G. Peano\rq\rq, University of Torino, via Carlo Alberto 10, I--10123 Torino, Italy}

\begin{abstract}
An ecosystem made of nutrients, plants, detritus and dissolved oxygen is presented.
Its equilibria are established.
Sufficient conditions for the existence of the coexistence equilibrium are derived and its
feasibility is discussed in every detail.
\end{abstract}


\end{frontmatter}


\section{Introduction}
Eutrophication is the large increase of aquatic plants due to the discharge in waterbodies
of chemical elements such as nitrogen and phosphorus.
Dead aquatic plants generate detritus, exploiting large amount of dissolved oxygen for its
decomposition, and therefore endangering aquatic life and fish survival.
We introduce here a mathematical model for such an ecosystem and investigate its behavior.

\section{Model formulation}
Let $N$, $D$ and $O$ respectively denote the nutrients, detritus and dissolved oxygen amount
in the water. Further, let $P$ denote the aquatic plants population.
The nutrients $N$ are assumed to be discharged into the lake
at rate $q$; they also are generated by the reconversion of decomposed detritus,
at rate $u$, a fraction $0<f<1$ of which becomes new nutrient.
This process involves also the usage of dissolved oxygen, and therefore
the term representing it must contain the product of both the required quantities,
$O$ and $D$.
In addition, nutrients can be washed away from the body water by emissaries, at rate $a$
and there are further losses due to nutrients uptake by the plants, that need them for
surviving. The uptake rate is denoted by $b$.
The plants reproduce because they assume nutrients, this grazing being expressed by
the product $NP$, but only a fraction $0<e<1$ of the assumed nutrients contribute to
increase the plant biomass. The plants have a natural mortality rate $m$, and experience
intraspecific competition at rate $c$.
The detritus originates by dead plants. A proportion $0<g<1$ of the plants that die sink
to the bottom of the water body and become detritus. In part the latter may be washed
away by outgoing streams, at rate $w$; in addition, using dissolved oxygen, detritus
at rate $u$ may be decomposed by microorganisms
into its basic elements, which contribute to the formation
of nutrients.
Finally, dissolved oxygen may be contained into the streams flowing into the lake, we assume
this to occur at rate $h$, but it is also produced by the aquatic plants
through photosynthesis, at rate $s$.
Because most of aquatic plants float on the surface of a lake, large amount of the
produced oxygen may go into the air, justifying the fact that the rate $s$ may be small.
In addition, there could be a shading effect, for which only the plants near the surface
get enough light, while the ones in the lower layers of water are in part shaded
from direct sunlight by the topmost ones.
In order for photosynthesis to be viable, plants need clorophyll and
consume nutrients for its production.
Thus, in the model we must model this process via the bilinear term $PN$.
Further, there is a natural upper limit on their oxygen production capacity, that produces
a saturation limit, a fact that is
expressed by using a Holling type II function to model it, $k$ being the half saturation
constant. In addition, dissolved oxygen is also washed away at rate $v$. Other losses are
due to the decomposition of detritus into its basic components, a process that we said
requires microorganisms that consume oxygen, at rate $z$.
With these assumptions, the model takes the following form:
\begin{eqnarray} \label{model1} 
 \frac{dN}{dt} = q+fuOD-aN-bNP,  \quad
 \frac{dP}{dt} = ebNP-mP-c P^2,  \\ \nonumber
 \frac{dD}{dt} = g(mP+cP^2)-wD-uOD, \quad
 \frac{dO}{dt} = h-vO+\frac{sPN}{k+PN}-zOD 
\end{eqnarray} 

\section{Boundedness of the system's trajectories}

We now evaluate the region of attraction for all solutions initiating in the positive octant.
Let $T=N+P+D+O$. Adding all the equilibrium equations, we obtain
\begin{eqnarray*}
\frac{d}{dt}T = q+h-aN-m(1-g)P -c(1-g)P^2 -b(1-e)NP -u(1-f) OD -z OD -wD -vO
+ s \frac {PN}{k+PN}.
\end{eqnarray*}
Dropping some of the negative terms, taking into account that $0\le e,f,g \le 1$ and
that the last fraction is smaller than one and defining
$\beta_m=min\{a,m(1-g),w,v\}$ and $H=q+h+s$, the above differential equations becomes
the following differential inequality 
\begin{eqnarray*}
\frac{d}{dt}T \leq q+h+s-aN-m(1-g)P-wD-vO \le H - \beta_m T.
\end{eqnarray*}
The solution of the associated differential equation is
$T=H\beta_m ^{-1} (1-\exp(-\beta_m t) + T(0) \exp (-\beta_m t)$.
Thus we obtain the final estimate that
defines the region of attraction for all solutions initiating in the interior
of the positive orthant:
$$
T(t) \le \max \{ H\beta_m ^{-1}, T(0)\}.
$$

\section{System's equilibria}
The model (\ref{model1}) has only two equilibria:
the environment containing only oxygen and nutrients, $E_3 =(qa^{-1}, 0, 0, hv^{-1})$,
and the system in which all
the components are at nonzero level $E^* =(N^*, P^*, D^*, O^*)$. 
In addition to the general case, there are several particular cases, that arise when
one or both of the exogenous inputs are blocked:
$E_0 =(0, 0, 0, 0)$ arising with $q=0$, $h=0$, $E_1 =(qa^{-1}, 0, 0, 0)$ whenever $h=0$,
$E_2 =(0, 0, 0, hv^{-1})$ obtained for $q=0$.

Feasibility of each $E_i$, $i=0,\ldots ,3$ is obvious. To analyze coexistence,
from the second equilibrium equation of (\ref{model1})
\begin{equation}\label{N*}
N^*=(m+cP)(eb)^{-1}\ge 0, \quad  NP=P(m+cP)(eb)^{-1}
\end{equation}
are obtained. Now substituting into
the first equilibrium equation, we find 
\begin{equation}\label{Psi}
OD = \Psi (P) (befu)^{-1}, \quad \Psi (P) \equiv bc P^2+(ac+bm)P+am-beq.
\end{equation}

The third equilibrium equation now gives
$D^* =[g(mP+cP^2)-uOD]w^{-1}$. Using (\ref{Psi}), we have explicitly:
\begin{eqnarray}\label{D*}
D^* = \frac{1}{befw} \Phi(P), \quad
\Phi(P) \equiv AP^2 + BP +C, \quad
A= bc\left( efg-1\right) <0, \quad
B= befgm-(ac+bm) = bm(efg-1)-ac <0.
\end{eqnarray}

\begin{eqnarray*}
C= beq - am.
\end{eqnarray*}
Now, 
the sign of $OD$ only depends on $\Psi(P)$. From (\ref{Psi}),
when $C>0$, (\ref{D*}), to guarantee $\Psi(P)\ge 0$, if $P_2$ is its positive root,
we need $P\geq P_2$.
When $C<0$ instead, $\Psi(P)\ge 0$ for every $P\ge 0$.
To impose $D\ge 0$, we study the concave parabola $\Phi(P)$, with roots $P_3\le P_4$.
For $C>0$, we have that $\Phi(P)\ge 0$ for $0 \le P\le P_4$,
going down from $(0,C)$ to $(P_4,0)$.
From (\ref{Psi}), (\ref{D*}):
\begin{equation}\label{O1*}
O_1(P)= \frac {OD}{D}
= \frac wu \frac {\Psi(P)}{\Phi(P)}.
\end{equation}
To guarantee its nonnegativity,
we combine the above results for $\Psi(P)$ and $\Phi(P)$, to get the
following condition and the corresponding nonnegativity interval:
\begin{equation}\label{OD}
(a)\ C>0, \quad B<0, \quad P_2 \leq P \leq P_4.
\end{equation}

Let $\chi (P)=behk+m(h+s)P+c(h+s)P^2 > 0$,
$\Pi (P)=bek+mP+cP^2 > 0$,
$\Gamma (P)=Q+MP+LP^2$, $Q=befvw+zC$, $M=zB<0$, $L=zA<0$ and
the discriminant of
$\Gamma$ be $\Delta_{\Gamma} = M^2-4LQ>M^2$.
Eliminating $NP$ from
(\ref{N*}),
the fourth equilibrium equation instead gives:
\begin{equation}\label{O2*}
O_2(P)=\left[ h+ sPN^* (k+P N^*)^{-1} \right] (v+zD^*)^{-1}
= befw \chi (P) [\Pi (P) \Gamma (P)]^{-1},
\end{equation}

To find sufficient conditions for the intersection of $O_1$ and $O_2$
for $P\ge 0$, we study each function separately.

$O_1$ is positive if conditions (\ref{OD}) hold. In the case
(a) $O_1$ increases from a zero at $P=P_2$ toward the vertical asymptote at $P=P_4$. 

The sign of $O_2$ is instead determined just by the one of $\Gamma$.
In principle, there is the following possible case guaranteeing
that the concave parabola $\Gamma$ is positive.
(1) $Q>0$, $M<0$ which implies a decreasing branch of this parabola lies in the first quadrant
between the vertical axis and its positive root $\gamma_+$, joining $(0,Q)$ with $(\gamma_+,0)$.

Correspondingly, there is a possible case for $O_2$ being nonnegative when $P\ge 0$:
it raises up from a positive height at the origin to the vertical asymptote at $P=\gamma_+$.

We now combine $O_1$ and $O_2$ to find an intersection in the first quadrant.
(1) is compatible with (a).
Sufficient conditions for $(P^*,O^*)$ to lie in the first quadrant is
\begin{eqnarray}\label{P*_suff}
(1)-(a): \quad Q>0, \quad B<0, \quad C>0, \quad P_2<P^*<P_4<\gamma_+.
\end{eqnarray}

\section{Stability}

Let $J$ be the Jacobian of (\ref{model1})
and let $J_i$ denote the matrix $M$ evaluated at each equilibrium $E_i$:
\begin{equation} 
J = \left(
\begin{array}{cccc}
-a-bP & -bN & fuO & fuD \\
ebP & ebN-m-2cP & 0 & 0 \\
0 & gm+2gcP  & -uO-w & -uD \\
\frac{skP}{(k+PN)^2} & \frac{skN}{(k+PN)^2} & -zO & -v-zD \\
\end{array}
\right)
\end{equation}
For $J_0$ the eigenvalues are all negative, namely $-a$, $-m$, $-w$, $-v$.
Therefore, $E_0$ is stable. $J_1$ has the following eigenvalues
$-a$, $(beq-am)a^{-1}$, $-w$, $-v$. From the condition (\ref{P*_suff}), $C = beq-am>0$ we can obtain that if the equilibrium $E^*$ exists, $E_1$ becomes a saddle point.
$J_2$ has four negative eigenvalues $-a$, $-m$, $-huv^{-1}-w$, $-v$ so that $E_2$ is
also always locally asymptotically stable.
The eigenvalues at the equilibrium $E_3$ are
$-a$, $(beq-am)a^{-1}$, $-hu v^{-1}-w$, $-v$ and again this equilibrium is a saddle point when $E^*$ exists.
The stability of the
coexistence point $E^*$ is examined numerically.

\section{Numerical simulations}

The equilibrium $E^*$ is feasible and stable for
the parameter values:
$q=0.5$, $f=0.5$, $u=0.5$, $a=0.05$, $b=0.41$, $e=0.9$, $m=.095$, $c=0.08$, $g=0.9$,
$w=0.013$, $h=0.3$, $v=0.08$, $s=0.02$, $k=0.02$, $z=0.025$,
obtaining
$N^*=0.7801$, $P^*=2.4107$, $D^*=0.3436$, $O^*=3.6098$.

\begin{figure}[ht]\label{q}
\begin{minipage}{40mm}
\includegraphics[width=4.cm]{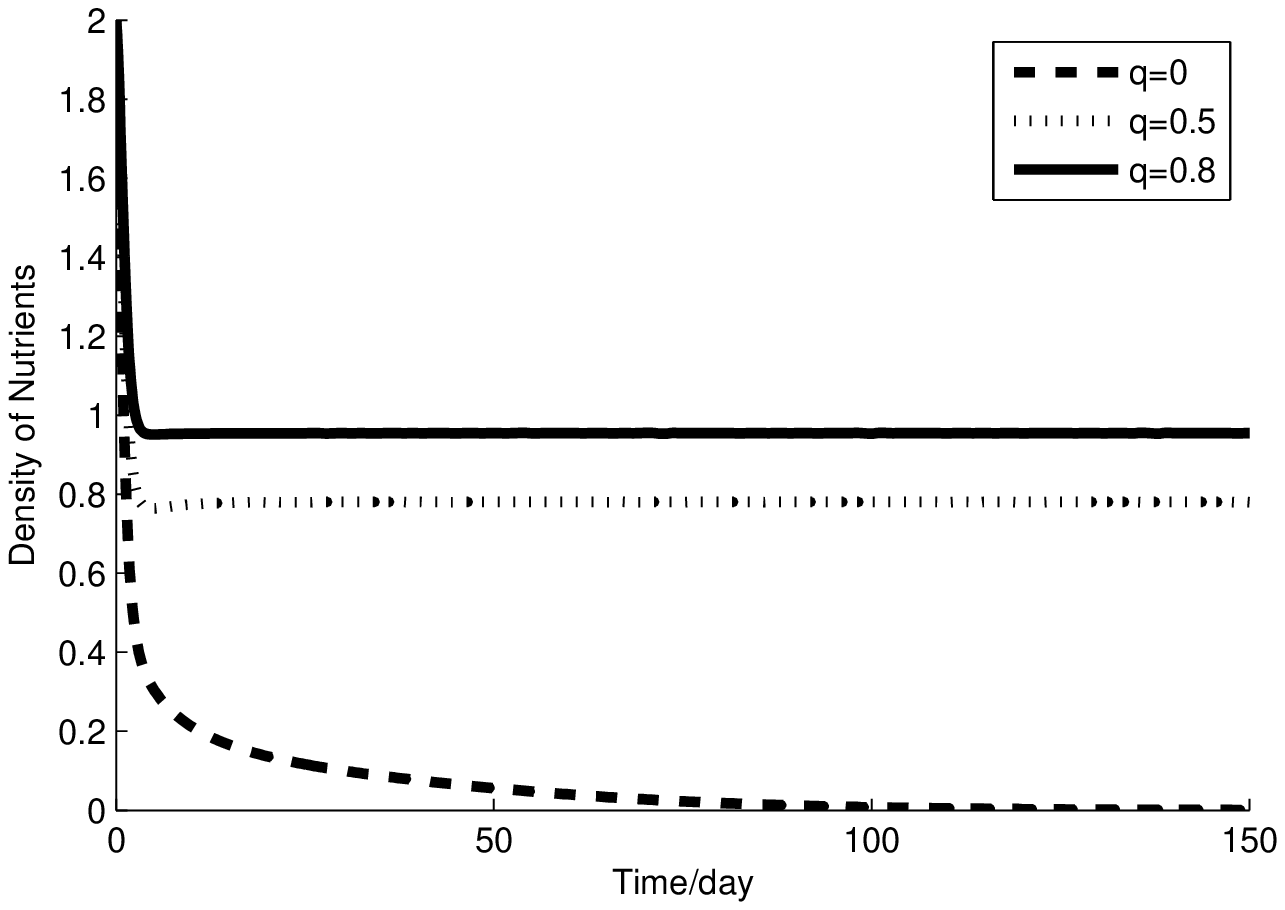}
\end{minipage}
\begin{minipage}{40mm}
\includegraphics[width=4.cm]{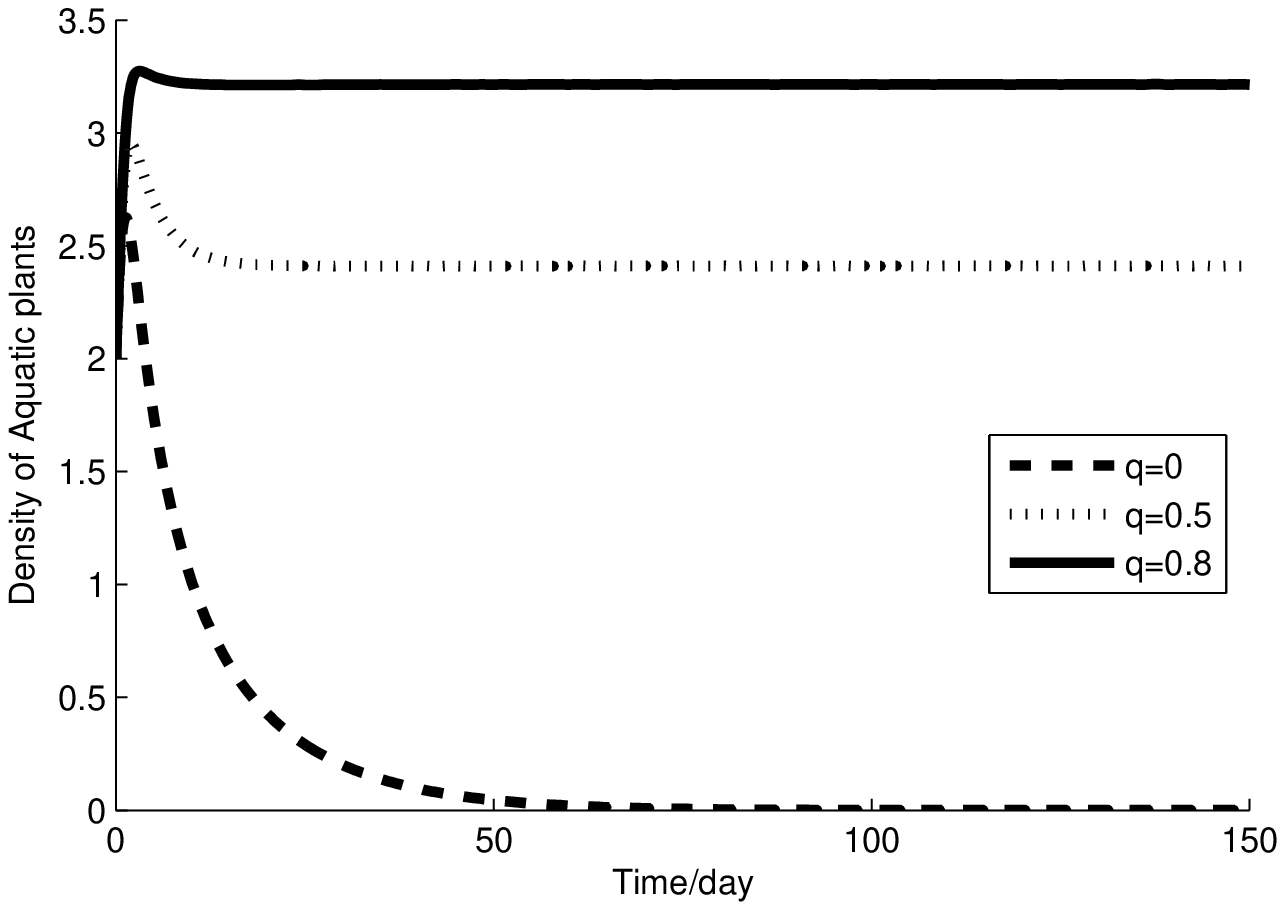}
\end{minipage}
\begin{minipage}{40mm}
\includegraphics[width=4.cm]{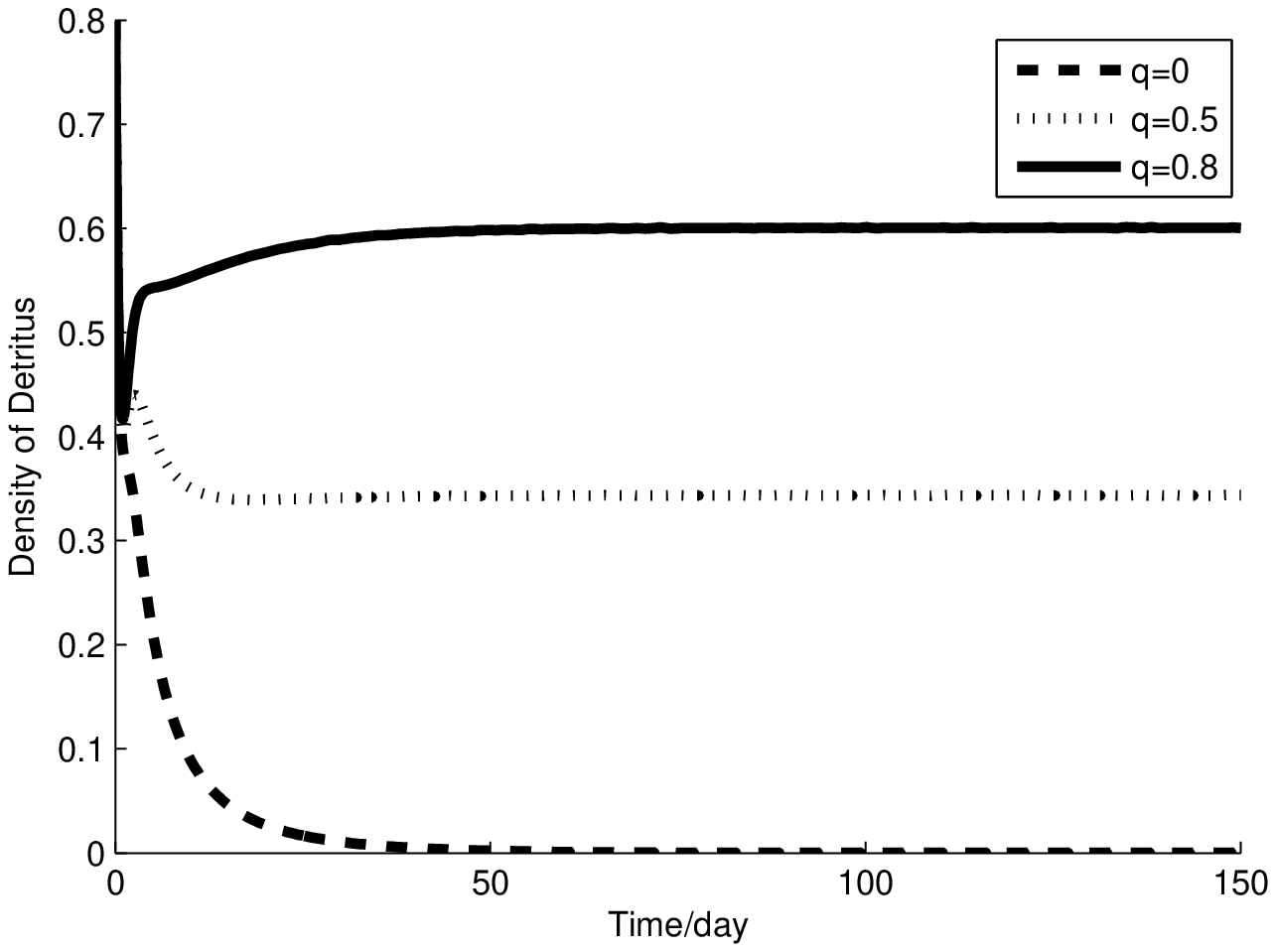}
\end{minipage}
\begin{minipage}{40mm}
\includegraphics[width=4.cm]{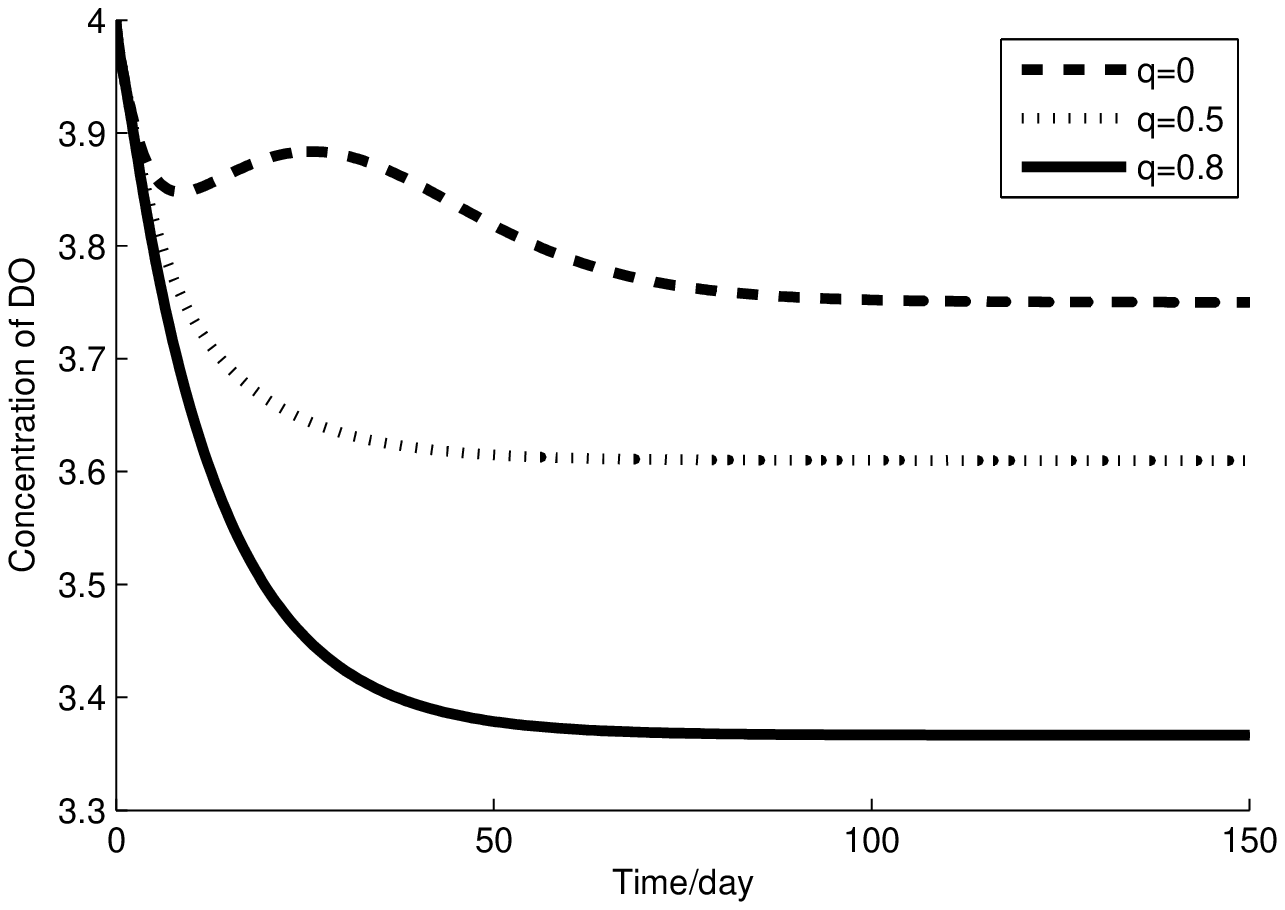}
\end{minipage}
\caption{Variations of nutrients, aquatic plants, detritus and dissolved oxygen
as functions of $q$. A reduction of the exogeous input of nutients lowers the equilibrium
values of all the ecosystem components but for dissolved oxygen.}
\end{figure}

Letting the input rate of nutrients vary, $q$,
from Figure \ref{q} we observe that with increasing $q$, the densities of nutrients,
aquatic plants and detritus increase.
In fact, for
$q=0$ their densities become zero within a short period.
The opposite results occur instead for dissolved oxygen, the concentration of which
increases with decreasing $q$. When $q=0$, dissolved oxygen attains its maximum concentration.

\begin{figure}[ht]\label{f}
\begin{minipage}{40mm}
\includegraphics[width=4.cm]{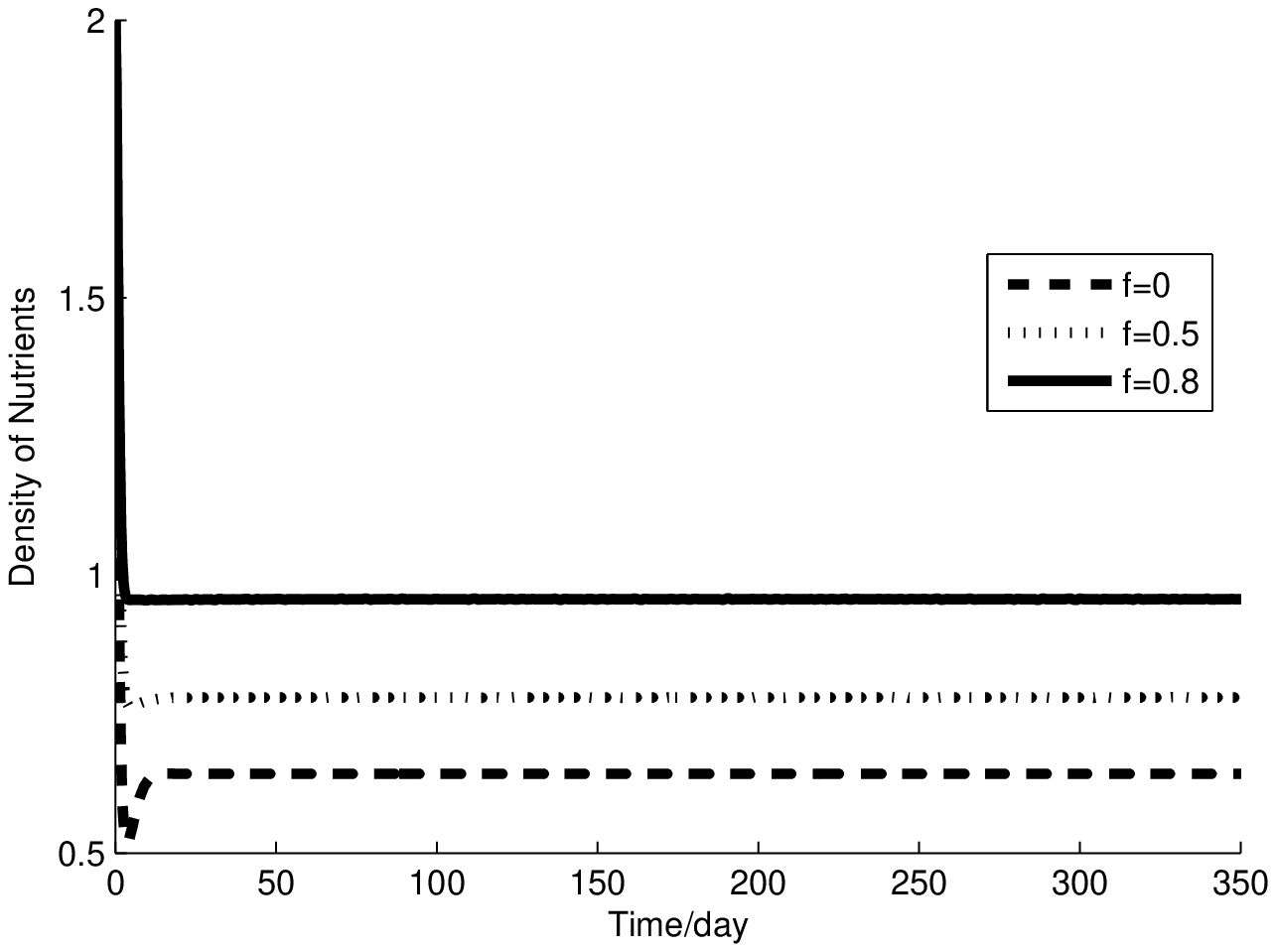}
\end{minipage}
\begin{minipage}{40mm}
\includegraphics[width=4.cm]{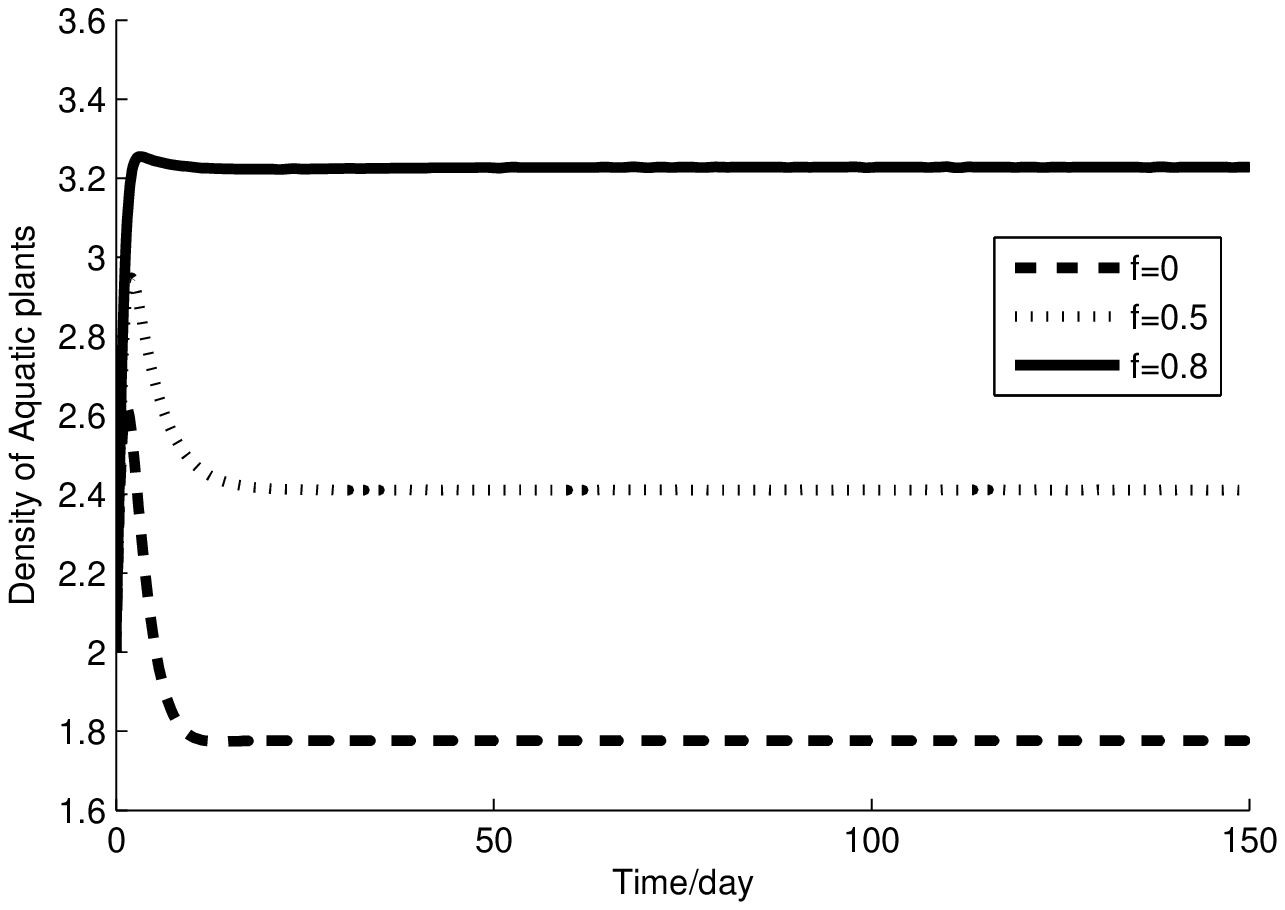}
\end{minipage}
\begin{minipage}{40mm}
\includegraphics[width=4.cm]{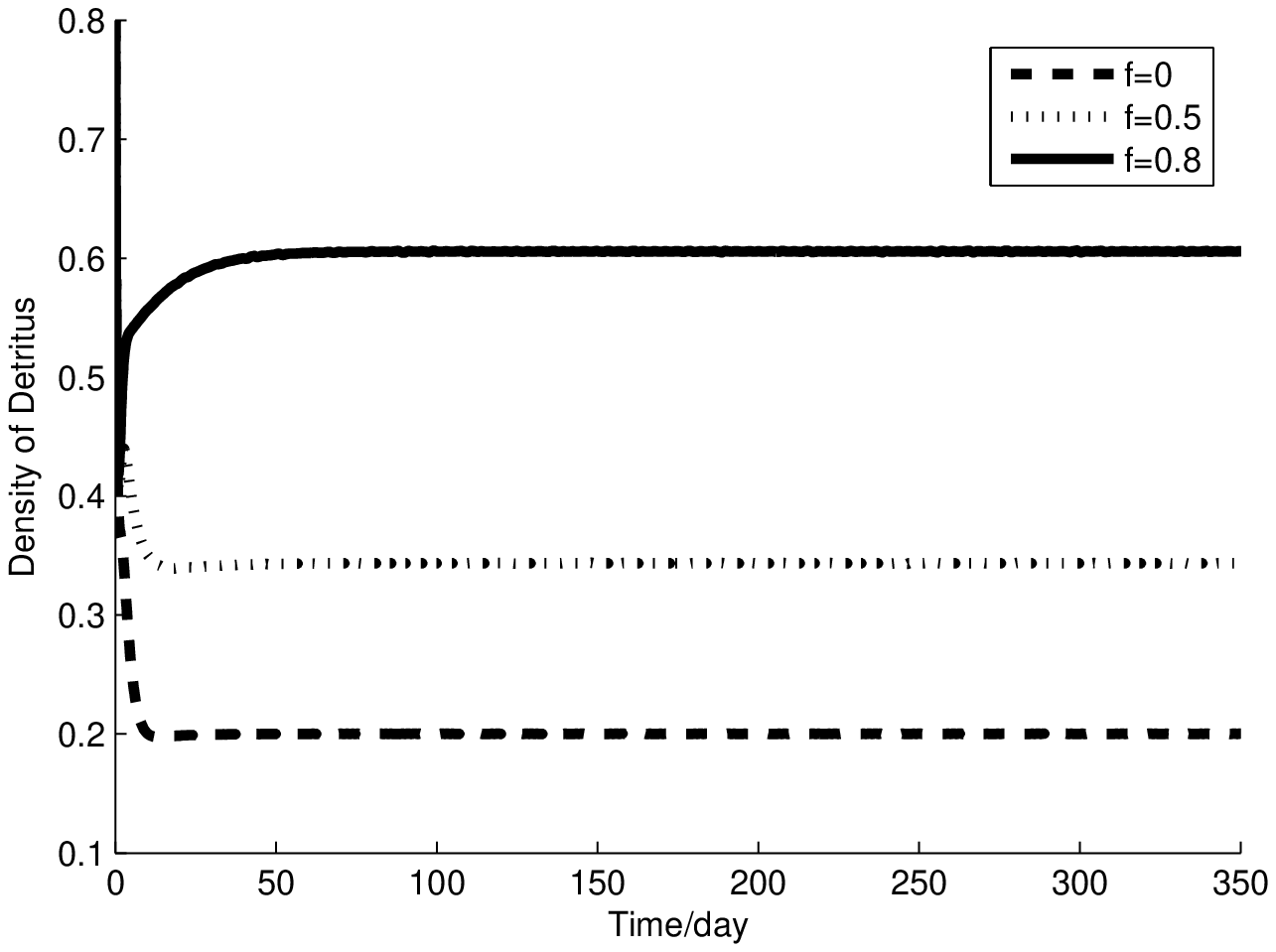}
\end{minipage}
\begin{minipage}{40mm}
\includegraphics[width=4.cm]{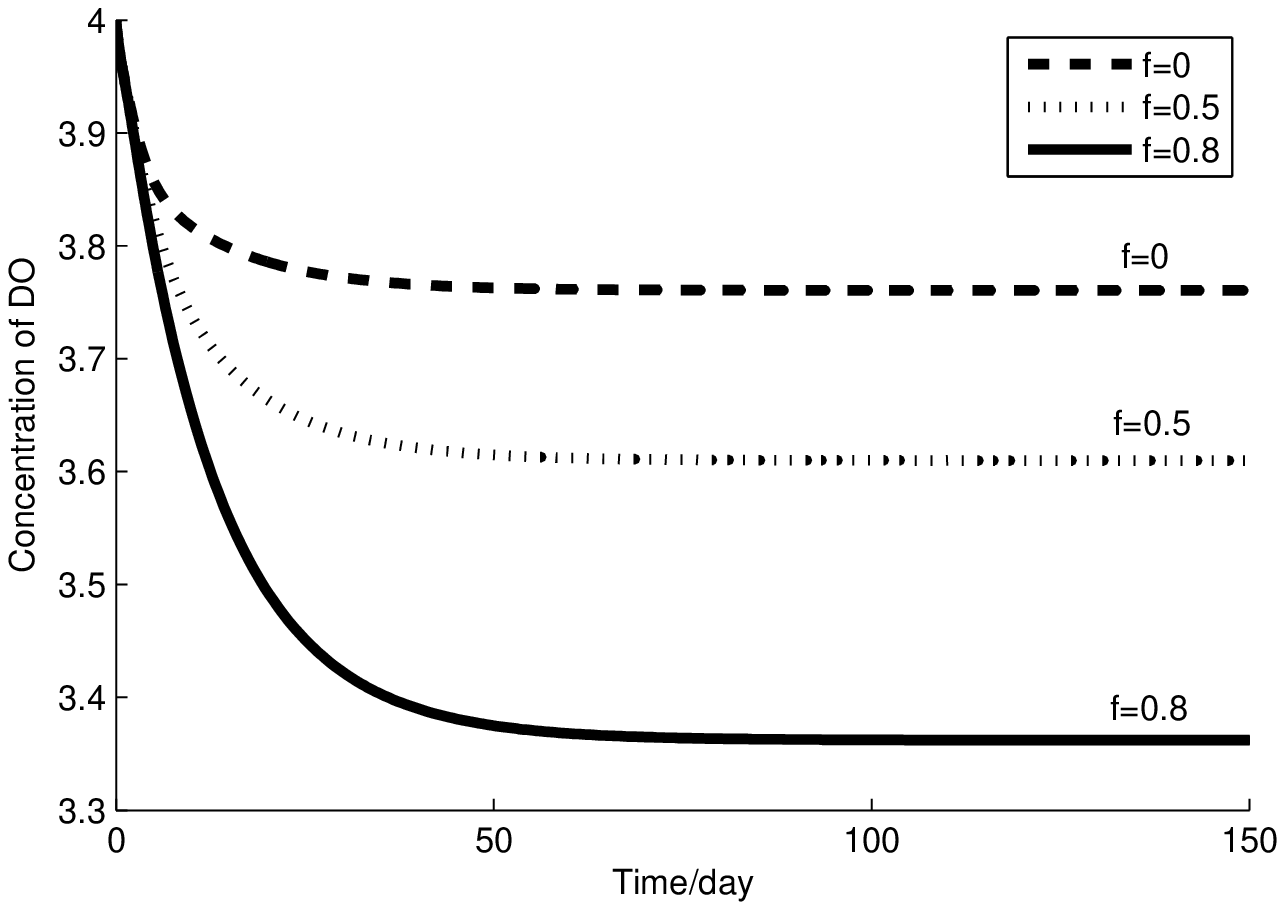}
\end{minipage}
\caption{Variations of nutrients, aquatic plants, detritus and dissolved oxygen
as functions of $f$. Reducing the conversion rate of detritus into nutrients
enhances dissolved oxygen but
hinders nutrients, aquatic plants and unexpectedly also detritus itself.
}
\end{figure}

Figure \ref{f} depicts instead the effect of rate of conversion of detritus into nutrients
on the coexistence equilibrium $E^*$. A similar situation as for Figure \ref{q} arises.
With increasing $f$, the densities of nutrients, aquatic plants and detritus increase,
while the concentration of dissolved oxygen decreases.

\section{Discussion}

The ecosystem could disappear, as the equilibrium $E_0$ is locally
asymptotically stable, not surprisingly. In fact a pond lacking sufficient
inputs becomes dry, and all its components disappear.
The no-life equilibrium
$E_2$ is also always stably achievable. $E_1$ and $E_3$ can be
rendered stable if $C<0$. Rewritten as $be m^{-1}<aq^{-1}$ it means that the reduced plants natality, i.e. the ratio of new plants over their mortality, should be bounded above by the reduced
nutrients depletion rate, i.e. the ratio of their washout rate to the net input rate.
Hence, more nutrients are supplied into the system than they are used for
plant reproduction. In such case the nutrients-only equilibrium and the point
with only nutrients and dissolved oxygen are stable. 

\vspace{-0.5cm}

\vspace{-0.5cm}

\begin{figure}[ht]\label{new}
\begin{minipage}{40mm}
\includegraphics[width=4.cm]{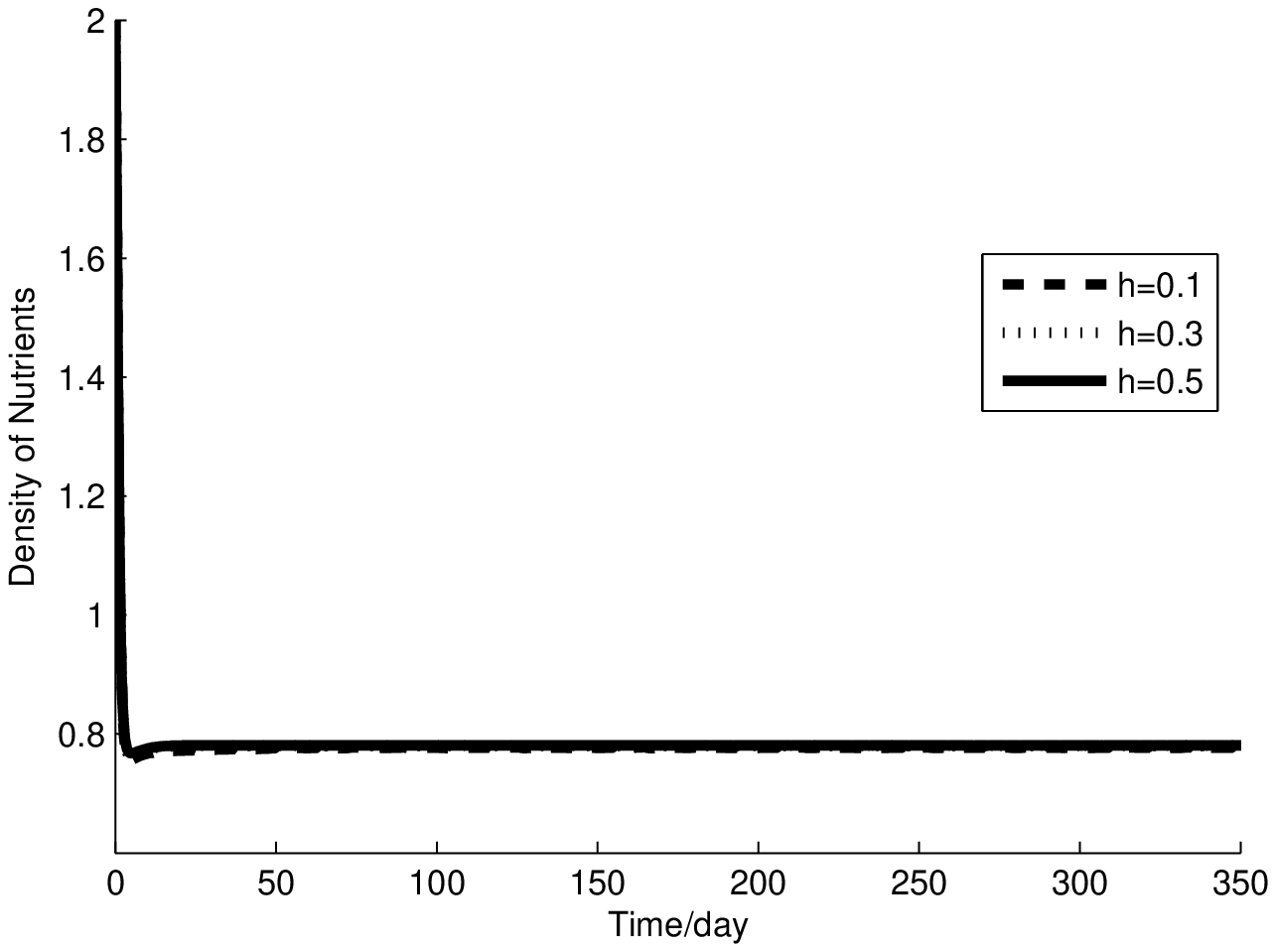}
\end{minipage}
\begin{minipage}{40mm}
\includegraphics[width=4.cm]{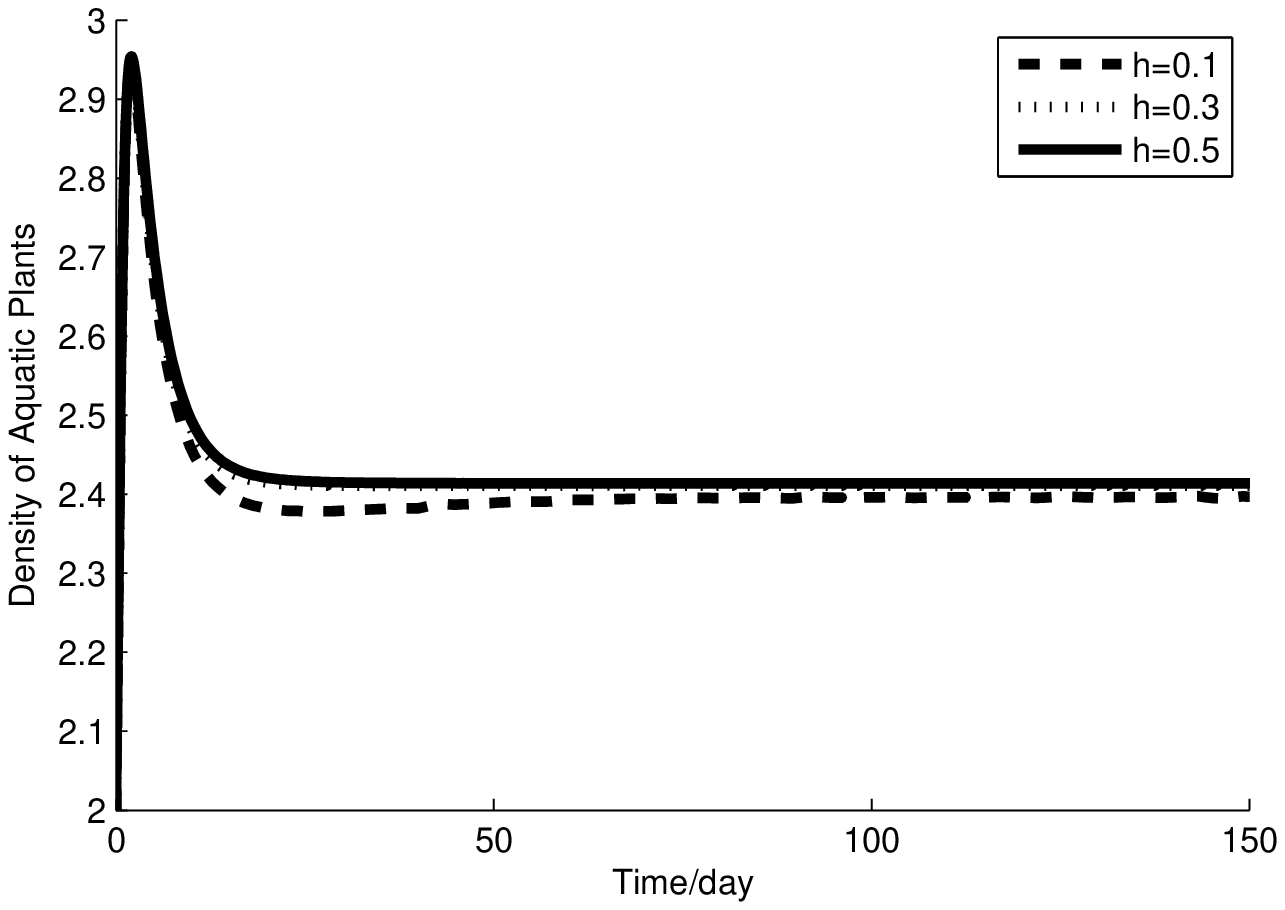}
\end{minipage}
\begin{minipage}{40mm}
\includegraphics[width=4.cm]{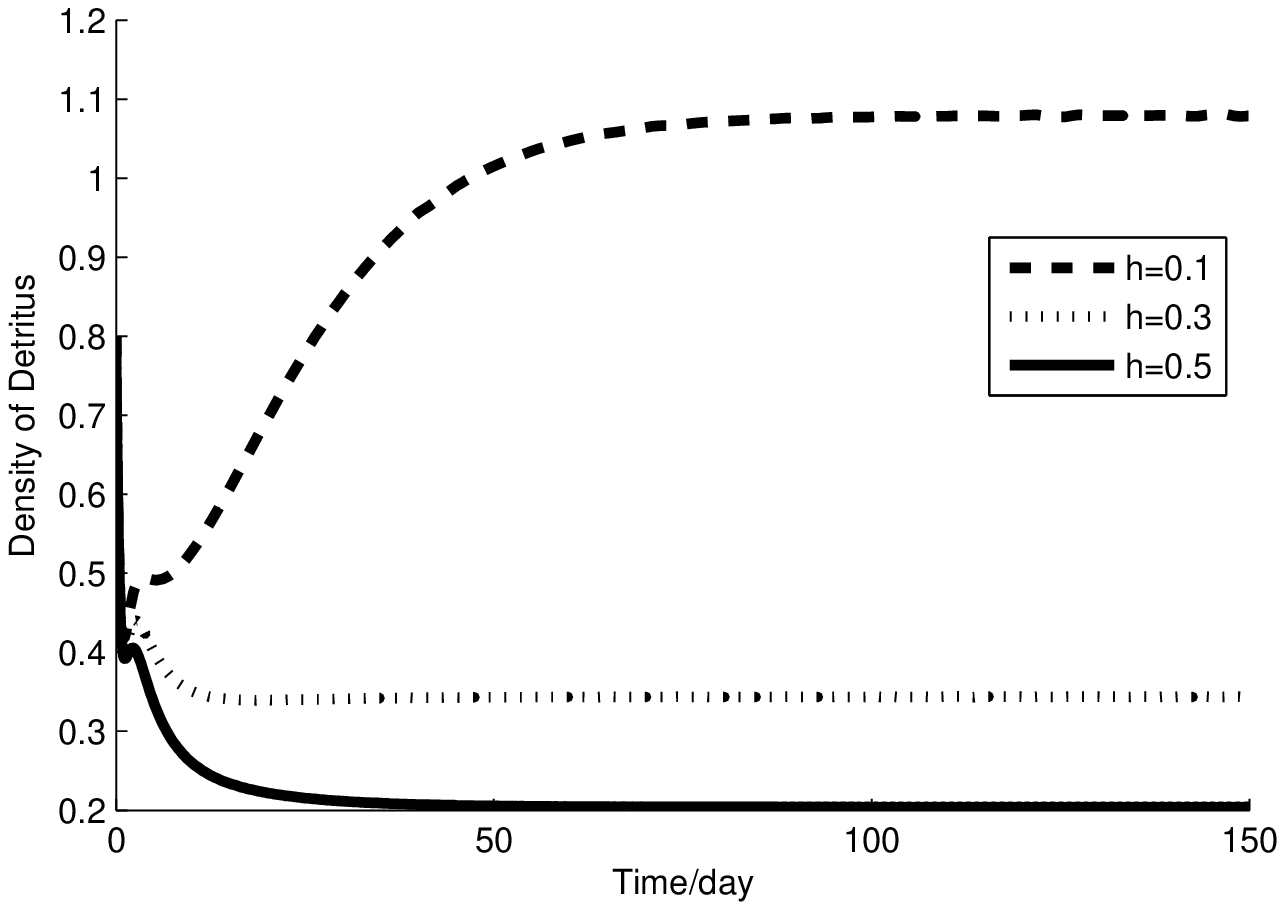}
\end{minipage}
\begin{minipage}{40mm}
\includegraphics[width=4.cm]{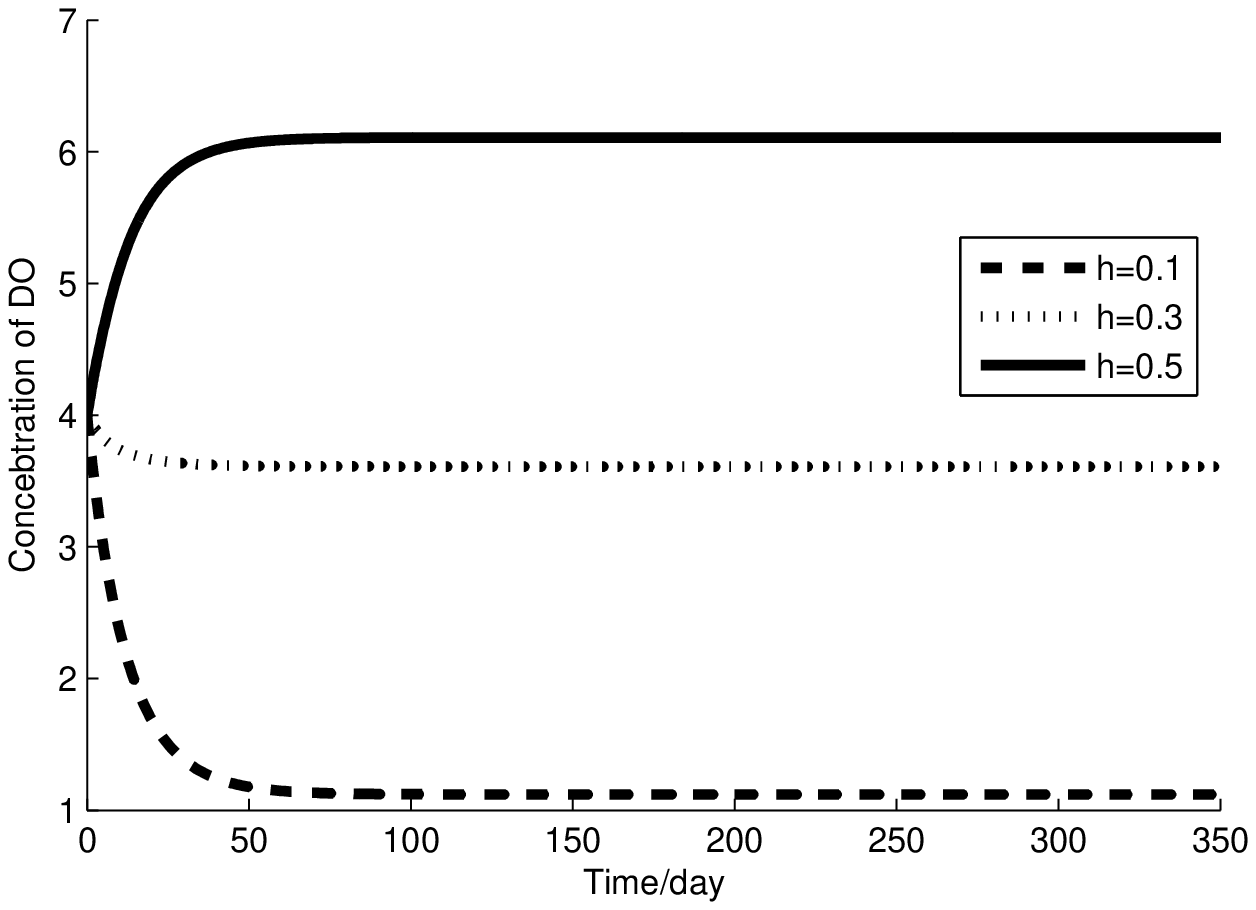}
\end{minipage}
\caption{Variations of nutrients, aquatic plants, detritus and dissolved oxygen
as functions of $h$. While nutrients and aquatic plants are scarcely affected by
changes in this parameter, for smaller values of $h$ we find a sharp increase in detritus and
a corresponding sensible decrease in the dissolved oxygen to very low concentrations.
To maintain an inflow of dissolved oxygen appears thus to be necessary to keep
the water body healthy.
}
\end{figure}

\section*{Acknowledgements}
This work has been partially supported by the projects
``Metodi numerici in teoria delle popolazioni'' and
``Metodi numerici nelle scienze applicate'' of the
Dipartimento di Matematica ``Giuseppe Peano'' of the Universit\`a di Torino.

\end{document}